\documentclass[conference]{IEEEtran}
\IEEEoverridecommandlockouts

\usepackage{amsmath,amssymb,amsfonts}
\usepackage{textcomp}
\usepackage[normalem]{ulem}

\usepackage{cite}
\def\BibTeX{{\rm B\kern-.05em{\sc i\kern-.025em b}\kern-.08em
    T\kern-.1667em\lower.7ex\hbox{E}\kern-.125emX}}

\usepackage[table]{xcolor}
\definecolor{distanceblue}{HTML}{6a97ec}
\definecolor{amd0}{HTML}{2F7B22}
\definecolor{amd1}{HTML}{59bc22}
\definecolor{amd2}{HTML}{77dd3b}
\definecolor{amd3}{HTML}{f2e41c}
\definecolor{amd4}{HTML}{fda403}
\definecolor{amd5}{HTML}{fe0d06}

\usepackage{algorithmic}
\usepackage{graphicx}
\usepackage{nicematrix}
\usepackage{booktabs}
\usepackage{multirow}
\usepackage{xfp}
\usepackage[font=small, skip=2pt]{caption}
\usepackage[export]{adjustbox}

\usepackage{enumitem}
\usepackage[hidelinks]{hyperref}
\usepackage{fancyhdr}
\fancypagestyle{ieeeheader}{
  \fancyhf{}
  \fancyfoot[C]{%
    \raisebox{0.8cm}{\noindent
      \begin{minipage}{\textwidth}
        \color{gray}\it\fontsize{6.7}{8.5}\selectfont
        \textcopyright\ 2026 IEEE. Personal use of this material is permitted. Permission from IEEE must be obtained for all other uses, in any current or future media, including reprinting/republishing this material for advertising or promotional purposes, creating new collective works, for resale or redistribution to servers or lists, or reuse of any copyrighted component of this work in other works.%
    \end{minipage}
    }
  }

}

\begin{document}

\title{Exploring the Latent Structure of AMD Severity Using Retinal Foundation Models\\\vspace{-.5em}
\thanks{Corresponding Author mail: panic.jovana.prime@osaka-u.ac.jp}
}

\author{%
  \IEEEauthorblockN{%
    Niccolò Fineschi\textsuperscript{1},
    Jovana Panic\textsuperscript{2,3},
    Atsushi Watanabe\textsuperscript{4},
    Hideaki Hayashi\textsuperscript{3},
    Kazuya Nishimura\textsuperscript{3},\\
    Yuki Maeno\textsuperscript{4},
    Samanta Rosati\textsuperscript{1},
    Gabriella Balestra\textsuperscript{1},
    Kohji Nishida\textsuperscript{2,4,5}, and
    Hajime Nagahara\textsuperscript{2,3}%
  }
  \IEEEauthorblockA{%
    \footnotesize
    \textsuperscript{1}Department of Electronics and Telecommunications, Polytechnic of Turin, Turin, Italy.\\
    \textsuperscript{2}Premium Research Institute for Human Metaverse Medicine (WPI-PRIMe), The University of Osaka, Osaka, Japan.\\
    \textsuperscript{3}D3 Center, The University of Osaka, Osaka, Japan.\\
    \textsuperscript{4}Department of Ophthalmology, Graduate School of Medicine, The University of Osaka, Osaka, Japan.\\
    \textsuperscript{5}Institute for Open and Transdisciplinary Research Initiatives (OTRI), The University of Osaka, Japan.\\
  }%
  \vspace{-3em}%
}

\maketitle
\thispagestyle{ieeeheader}

\begin{abstract}
Retinal foundation models have demonstrated strong performance across a wide range of ophthalmic tasks, yet relatively little is known about how disease-related information is represented within their learned embeddings and whether this distribution influences downstream prediction performance.
In this work, we use Age-related Macular Degeneration (AMD) as a case study to investigate how disease severity is represented within the embedding space of RET-CLIP, a retinal vision-language foundation model, using bilateral fundus images from the AREDS dataset, benchmarked against non-retinal encoders.
We first characterize the distribution of AMD severity within the learned representation space and then evaluate how this structure influences classification performance across three clinically relevant task formulations of increasing granularity. 
Finally, we analyze the most influential latent dimensions combining feature attribution, statistical phenotype association, and qualitative image analyses. 
Our results show that RET-CLIP represents AMD severity as a continuous spectrum rather than a set of discrete categories, with healthy and advanced disease stages exhibiting clearer separation than intermediate stages.
This organization was directly reflected in downstream performance, which decreased from a balanced accuracy of 0.844 and 0.814 for binary AMD and Late-AMD detection respectively to 0.446 for six-class grading.
Furthermore, the most influential latent dimensions were statistically associated to distinct retinal phenotypes, including drusen accumulation, pigmentary alterations, and advanced retinal lesions. 
These findings suggest that the structure of retinal foundation-model representations influences the difficulty of clinically relevant prediction tasks and demonstrate the potential of representation-level analysis for understanding disease organization, improving interpretability, and supporting future biomarker discovery.
\end{abstract}


\vspace{-1.3em}
\section{Introduction}

\begin{figure*}[!ht] 
  \centering 
  \includegraphics[width=\linewidth]{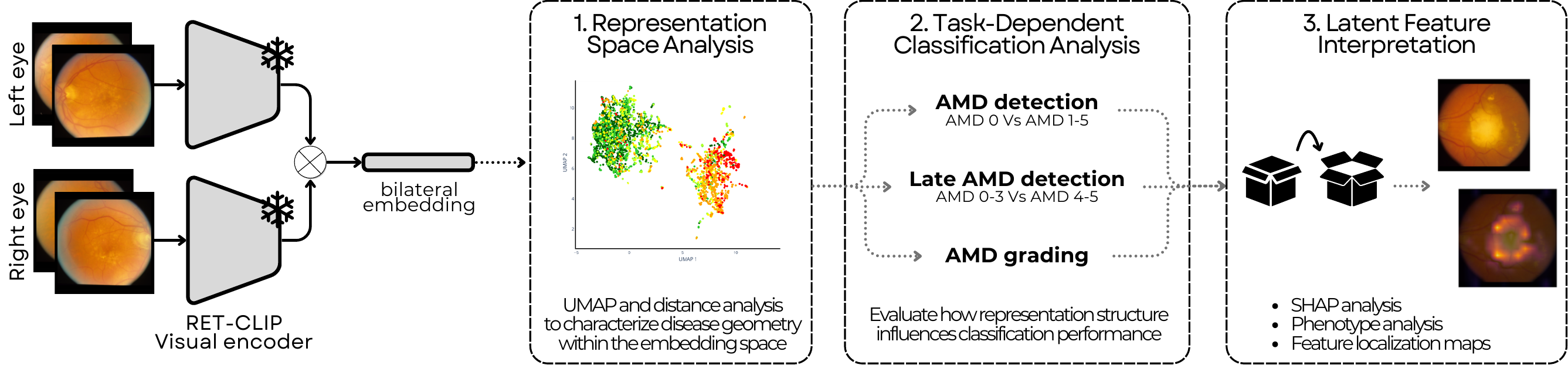} 
  \caption{Overview of the proposed analysis pipeline. RET-CLIP bilateral embeddings are investigated through latent-space analysis, task-dependent AMD classification, and latent feature interpretation.} 
  \label{fig:flow}
  \vspace{-1.5em}
\end{figure*}

Age-related Macular Degeneration (AMD) is a progressive retinal disease and one of the leading causes of irreversible vision loss worldwide \cite{vyawahare2022age}.
Disease severity ranges from healthy retinas to advanced stages characterized by AMD-related abnormalities such as drusen, pigmentary alterations, geographic atrophy, and macular neovascularization \cite{guymer_age-related_2023}.
To support risk stratification and monitoring of disease progression, the \textit{Age-Related Eye Disease Study} (AREDS), a longitudinal prospective study, introduced AREDS Simplified Severity Scale, a standardized patient-level score ranging from 0 to 5, based on findings from both eyes \cite{agron2024updated}. Accurate severity assessment is clinically important for guiding therapeutic decisions and identifying patients at increased risk of progression \cite{vyawahare2022age}.

Despite the availability of standardized grading systems, translating AMD severity assessment into a well-defined machine learning problem remains challenging.
Indeed, AMD progression is inherently continuous, while current automated systems typically require discrete prediction targets.
As a result, AMD has been investigated through different task formulations, ranging from binary disease detection \cite{peng2019deepseenet, lee2025artificial} to late-stage identification \cite{skevas2022simultaneous, yan2020deep, ganjdanesh2022longl, yin2022predicting, peng2020predicting, heo2020development, el2023scale} and fine-grained severity grading \cite{xie_automation_2023, peng2019deepseenet, agron2024updated}. 
While promising results have been reported, relatively little attention has been paid to understanding how AMD-related information is organized within the learned representation space and whether this distribution supports the different task formulations adopted in the literature. 

Retinal foundation models provide a unique opportunity to investigate this question. 
Trained on large collections of retinal fundus images paired with clinical reports, models such as RET-CLIP \cite{du2024ret} acquire rich visual representations that transfer effectively across multiple ophthalmic tasks. 
Although retinal foundation models have demonstrated strong downstream performance, it remains unclear how disease-related information is organized within their embedding spaces, which latent dimensions drive clinical predictions, and how these learned representations influence the difficulty of clinically relevant prediction tasks.
Understanding such internal representation structures has recently attracted increasing attention in foundation-model interpretability research \cite{jiao2025foundation}.

AMD represents an ideal case study for investigating retinal foundation-model representations.
Its continuous progression, clinically challenging intermediate stages, and well-established severity scale make it possible to examine whether the assessment difficulty arises primarily from the classification task itself or the structure of the underlying disease representation. 

This study aims to provide new insights into how retinal foundation models encode disease-related information and how the organization of learned representations influences clinically relevant prediction tasks, ultimately supporting more interpretable systems and future biomarker discovery strategies. 
In particular, we investigate how AMD severity is represented within RET-CLIP embeddings using bilateral retinal fundus images from the AREDS dataset.
Using visual embeddings extracted from the RET-CLIP encoder, we first analyze how AMD severity is organized within the learned representation space.
We then assess how the underlying representation influences classification performance across three clinically relevant task formulations of increasing granularity: (i) healthy versus AMD (AREDS score 0 versus scores 1--5), (ii) non-late versus late AMD (scores 0--3 versus scores 4--5), and (iii) full six-class AMD grading (scores 0--5). 
Finally, we combine SHAP-based feature attribution with image-level interpretation to investigate which retinal phenotypes are encoded within the most influential latent dimensions.

\section{Materials and Methods}

\subsection{Dataset and Pre-processing}
We used data from the \textit{Age-Related Eye Disease Study} \cite{agron2024updated}, including bilateral retinal fundus images, AREDS Simplified Severity Scale score and clinical information across multiple follow-up visits.
From the full cohort, \(448\) patients were retained after excluding \(170\) subjects with identifier mismatches. Since our analysis was based on bilateral retinal information, visits with incomplete image pairs (i.e., missing left or right eye acquisitions) and missing AMD severity labels were discarded, resulting in a final dataset of \(3931\) retinal fundus eye-couple images.
Since no image quality assessment was performed, the dataset preserves the unavoidable variability of clinical acquisitions and may include images affected by acquisition artifacts.
Moreover, all retinal fundus images were preprocessed by cropping the retinal region using Otsu thresholding, zero-padding to obtain a square aspect ratio, resizing to $224 \times 224$ pixels, and applying the default RET-CLIP normalization procedure.

\subsection{RET-CLIP Visual Embeddings}\label{sub:retclip}
For each retinal image, feature embeddings were extracted using the frozen RET-CLIP visual encoder (Fig.\ref{fig:flow}). 
Independent \(512\)-dimensional feature vectors from left and right eyes were concatenated to obtain a single \(1024\)-dimensional patient-level representation for the following experiments.

\subsection{Experiments}
The overall study design is shown in Fig.~\ref{fig:flow}. To investigate how AMD severity is represented within RET-CLIP embeddings, we carried out three experiments: (1) representation-space analysis, (2) task-dependent classification analysis, and (3) most influential latent features interpretation.
Because the AREDS dataset is longitudinal, individual visits serve as the unit of observation.
To account for within-patient dependence and data leakage, resampling during data splitting and uncertainty quantification was performed at the patient level.

\subsubsection{Representation Space Analysis}\label{sub:Experiment1}
To investigate how AMD severity is represented within the RET-CLIP embedding space, we visualized the \(1024\)-dimensional bilateral embeddings using Uniform Manifold Approximation and Projection (UMAP) \cite{mcinnes2018umap} across patient-level AMD severity scores (0--5), as well as the presence of drusen and pigmentary abnormalities, geographic atrophy, and macular neovascularization, typically marking the late-AMD status \cite{guymer_age-related_2023}  (Fig.~\ref{fig:flow}.1).
To quantify class separation and overlap on the full \(1024\)-dimensional space we evaluated centroid cosine distances (inter-class separation), mean pairwise cosine distances (intra-class variability), and a Fisher-like discriminability ratio (between-class distance normalized by within-class scatter).
Uncertainty across all measures was estimated via patient-cluster bootstrap and reported as 95\% confidence intervals.

\subsubsection{Task-dependent Classification}\label{sub:Experiment2}
To investigate how the organization of the RET-CLIP representation space influences downstream AMD assessment, we evaluated three clinically relevant classification tasks corresponding to increasing levels of disease granularity  (Fig.~\ref{fig:flow}.2):
\begin{enumerate}[label=\alph*)]
    \item \textbf{AMD detection}: ``Healthy/No AMD`` (AMD 0) vs. ``AMD`` (scores 1--5).
    \item \textbf{Late-AMD detection}: ``Non-late AMD`` (AMD 0--3) vs. ``Late AMD`` (AMD 4--5).
    \item \textbf{AMD grading}: Multi-class classification into six severity levels (scores 0--5).
\end{enumerate}
These tasks were selected to progressively increase the granularity of disease assessment, allowing us to investigate whether the learned representation supports coarse disease discrimination and fine-grained severity grading to the same extent.

For model development, the dataset was first divided into construction and test sets using an \(80\%-20\%\) partition, and the construction set was then further divided into training and validation subsets using the same \(80\%-20\%\) partitioning.
To prevent data leakage, all bilateral image pairs from the same patient were assigned to a single subset. 
No class-based stratification was applied, preserving the original distribution of AMD severity stages.

For each task, we evaluated multiple machine learning classifiers, including AdaBoost, Multi-Layer Perceptrons (MLP), Random Forest (RF), and Support Vector Machines (SVM). 
By keeping the RET-CLIP embeddings fixed and varying only the classifier, we assessed whether the observed performance trends were consistent across different learning paradigms.
Model robustness was evaluated by repeating the training procedure on ten independent random partitions of the construction set while maintaining the same held-out test set.
Performance was summarized as mean and standard deviation. 

Overall classification performance was evaluated using balanced accuracy (balacc) and overall accuracy (acc) for all three tasks.
For the selected classifiers, balacc was additionally summarized with 95\% confidence intervals from patient-cluster bootstrap replicates.
For the binary classification tasks, sensitivity (sens), specificity (spec), and the area under the receiver operating characteristic curve (AUC) were additionally reported.
For the multi-class AMD severity grading task, the mean absolute error (MAE) was used to account for the ordinal nature of the AMD severity scores.
Statistical differences between model performances within the same task formulation were assessed using the paired patient-cluster bootstrap, and a 95\% CI of the paired difference excluding zero adjusted with Benjamini--Hochberg false discovery rate (BH-FDR) was interpreted as evidence of a significant difference.

To evaluate the contribution of domain-specific pretraining, general-domain foundation model embeddings, from the CLIP visual encoder and an ImageNet-pretrained ResNet-50, were evaluated under the same predictive framework as controlled baselines.
This allowed us to assess whether performance differences were primarily driven by the clinical task formulation and the structure of the learned representation space, rather than by a particular classification algorithm.

\subsubsection{Latent Feature Interpretation}\label{sub:Experiment3}
To investigate which retinal phenotypes encoded within the RET-CLIP representation space contributed to AMD severity assessment, we performed a feature-level interpretability analysis combining feature attribution, statistical phenotype association, and qualitative image inspection (Fig.~\ref{fig:flow}.3). 
For each task formulation, SHapley Additive exPlanations (SHAP) \cite{lundberg2017unified} was applied to the best-performing validation set-wise classifier across all repeated training runs.
SHAP values were also evaluated on the validation sets and aggregated to obtain a global ranking of latent feature importance, and the three latent dimensions that were consistently identified across repetitions were selected for further analysis on the hold-out test set.

To quantitatively characterize the semantic content encoded by the identified latent dimensions, statistical association analyses were performed between feature activations and the clinical annotations available in the AREDS dataset.
Since the metadata included both ordinal and binary clinical variables, the statistical tests were selected according to the measurement scale of each phenotype under investigation.
Associations with ordinal variables, including the AMD severity score, were assessed using Spearman's rank correlation test, whereas binary phenotypes were evaluated using the Brunner--Munzel test.
To account for multiple comparisons, \textit{p-values} were adjusted using BH-FDR. 
For statistically significant associations after correction, association strength was quantified using Spearman's rank correlation coefficient for ordinal phenotypes, and the Brunner--Munzel rank-biserial correlation for binary ones.

To complement the quantitative analysis, representative retinal images exhibiting the highest feature activations from the test set were retrieved for qualitative inspection.
In addition, latent-dimension localization maps were generated by weighting patch-level activations from the RET-CLIP visual encoder with the attention scores between the transformer classification token and individual image patches, and projected back onto the image space.
Unlike GradCAM-based approaches, these maps do not explain the classifier prediction, but localize the retinal regions in which the feature is most strongly expressed. 

Since bilateral representations were obtained by concatenating features extracted independently from each eye, feature interpretation was performed at the individual-eye level.
The resulting observations obtained by the statistical phenotype analysis, representative high-activation images, and localization maps should be interpreted as an exploratory framework of the semantic content encoded by RET-CLIP rather than a direct visualization of the classifier features.

\subsection{Implementation Details}
All analyses were carried out on Python, using the PyTorch and scikit-learn libraries, on one Intel® Xeon® Silver 4510 processor with 64GB of RAM.
Detailed implementation settings, including preprocessing, UMAP configuration, classifier hyperparameters, SHAP analysis, and statistical procedures, together with the analysis code, are provided in the accompanying GitHub repository: \href{https://github.com/niccofine/AMD-Latent}{github.com/niccofine/AMD-Latent}.

\begin{figure*}[!ht] 
  \centering 
  \includegraphics[width=\linewidth]{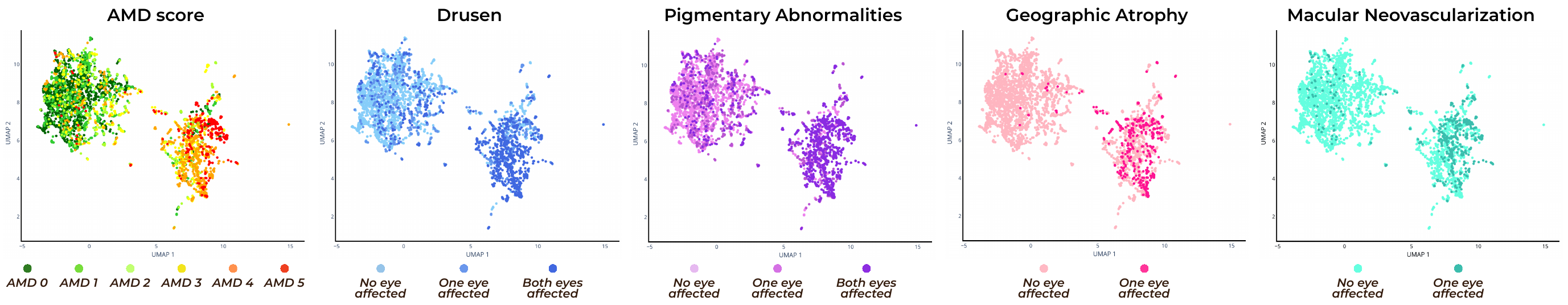} 
  \caption{UMAP visualization of RET-CLIP bilateral embeddings colored according to AMD severity and associated retinal biomarkers (drusen, pigmentary abnormalities, geographic atrophy, and macular neovascularization respectively)} 
  \label{fig:umap} 
  \vspace{-1.2em}
\end{figure*}

\section{Results and Discussion}

\subsection{Representation Space Analysis}

\begin{figure}[ht!] 
  \centering
  \includegraphics[width=\linewidth]{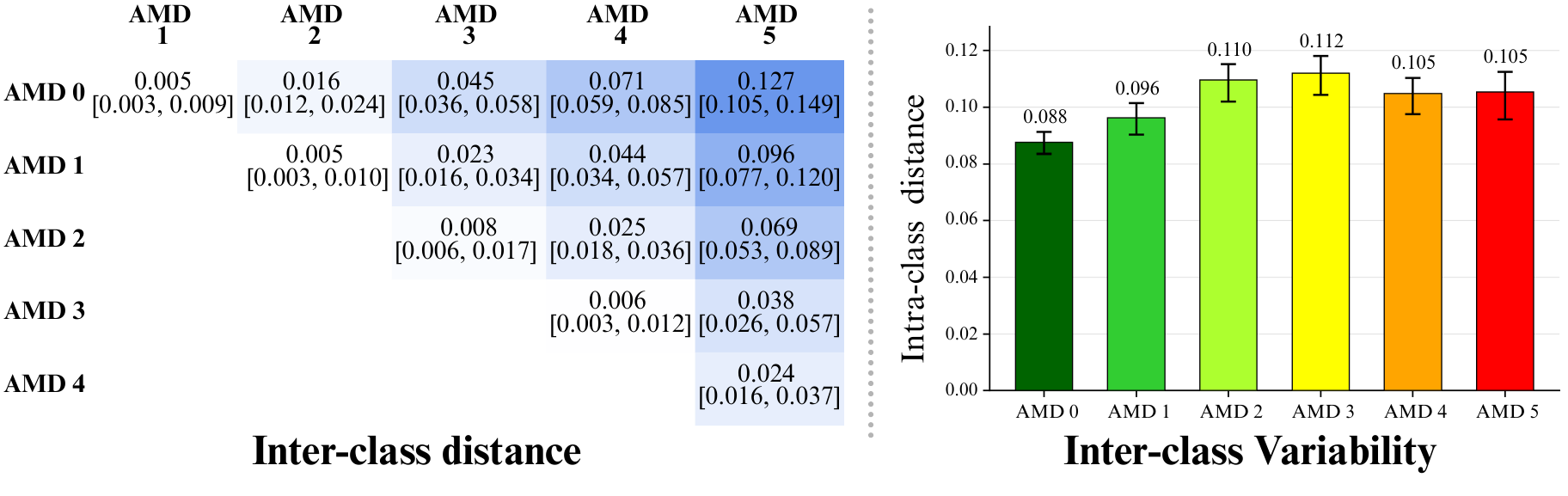}%
  \caption{Pairwise inter-class separation (left) and intra-class variability (right) between AMD severity groups, with 95\% confidence intervals from patient-cluster bootstrap. Darker cells indicate larger inter-class distances.}
  \label{fig:distances}
  \vspace{-2em}
\end{figure}

Fig. \ref{fig:umap} presents the UMAP projection of the RET-CLIP bilateral embeddings, while Fig. \ref{fig:distances} the quantitative analyses of inter-class distances and intra-class variability.
Rather than forming six clearly separated clusters, the AMD severity scores follow a continuous, overlapping gradient from healthy to advanced disease.
Healthy cases (AMD 0) and advanced disease (AMD 4--5) occupy distinct regions of the representation space, whereas intermediate stages exhibit substantial overlap and gradually bridge the transition between the two extremes.
This behavior becomes particularly evident when the latent space is colored according to retinal biomarkers underlying the AREDS grading system.
Both drusen and pigmentary abnormalities follow the same global organization observed for AMD severity, with unaffected eyes concentrated in the healthy region and bilateral abnormalities predominantly located in the advanced disease region.
A similar pattern was observed for geographic atrophy and macular neovascularization, associated with late stages of the disease, which were largely confined to the region occupied by AMD scores 4 and 5.
These findings suggest that RET-CLIP encodes AMD-related retinal characteristics within a coherent representation structure, in which disease severity emerges as a gradual progression rather than a set of isolated categories.

The inter-class distance analysis further supports this observation (Fig. \ref{fig:distances}), with distances between AMD categories increase progressively with disease severity: neighboring stages remain close (e.g., AMD~0 vs.~1: 0.005 [95\% CI: 0.003, 0.009]; AMD~4 vs.~5: 0.024 [0.016, 0.037]), while distant stages separate substantially more, reaching a maximum between AMD~0 and AMD~5 (0.127 [0.105, 0.149]).
These findings are further confirmed by Fisher-like discriminability ratio, which increases from 0.000 [0.000, 0.001] for AREDS~3--4 to 0.083 [0.056,0.120] for AREDS~0--5.
Together, these trends indicate that the latent representation reflects the gradual AMD progression rather than discrete categories. 
At the same time, intra-class variability remains relatively high across all stages and reaches its maximum for AMD score 3.
This finding is consistent with the known clinical heterogeneity of intermediate AMD stages and with reported reproducibility limitations in clinical AMD severity grading \cite{peng2019deepseenet}.
The high variability highlights the intrinsic complexity of fine-grained AMD assessment and suggests that intermediate stages occupy wider and overlapping regions of the representation space.

Overall, these results indicate that RET-CLIP encodes AMD-related information in a structured and clinically meaningful manner.
The gradual organization of disease severity, together with the high variability observed in intermediate stages, closely reflects current clinical understanding of AMD progression.
This suggests that retinal foundation-model embeddings capture relevant disease characteristics, preserving the inherent complexity of the AMD severity spectrum.

\subsection{Task-dependent Classification}
Table \ref{tab:dataset} reports the distribution of AMD severity scores within each partition and for each task.
Across the dataset, AMD severity stages were unevenly distributed, with AMD scores 0 and 4 representing the largest groups, while AMD scores 2, 3, and 5 were comparatively underrepresented.
Consequently, both binary tasks exhibited class imbalance, particularly Late-AMD detection, where late-stage cases accounted for approximately one quarter of the samples.

\begin{table}[t!]
    \centering
    \caption{Dataset splits and class distribution in terms of eye-couples (construction and test set in \textbf{bold} and \textit{italic}, respectively)}
    \label{tab:dataset}
    \renewcommand{\arraystretch}{1.3}
    \resizebox{\linewidth}{!}{%
    \begin{tabular}{lcccccc}
        \toprule
        & \textbf{AMD 0} & \textbf{AMD 1} & \textbf{AMD 2} & \textbf{AMD 3} & \textbf{AMD 4} & \textbf{AMD 5} \\
        \midrule
        \textbf{Total} 
        & 1387 & 637 & 418 & 456 & 710 & 323 \\

        \addlinespace[0.2em]

        \textbf{AMD} 
        & \cellcolor{green!20}\textbf{1161} / \textit{226} 
        & \multicolumn{5}{c}{\cellcolor{red!20}\textbf{2142} / \textit{402}} \\

        \addlinespace[0.2em]

        \textbf{LateAMD} 
        & \multicolumn{4}{c}{\cellcolor{green!20}\textbf{2446} / \textit{452}} 
        & \multicolumn{2}{c}{\cellcolor{red!20}\textbf{857} / \textit{176}} \\

        \addlinespace[0.2em]

        \textbf{AMD score} 
        & \cellcolor{amd0!70}\textbf{1161} / \textit{226} 
        & \cellcolor{amd1!70}\textbf{547} / \textit{90} 
        & \cellcolor{amd2!70}\textbf{364} / \textit{54} 
        & \cellcolor{amd3!70}\textbf{374} / \textit{82} 
        & \cellcolor{amd4!70}\textbf{584} / \textit{126} 
        & \cellcolor{amd5!70}\textbf{273} / \textit{50} \\
        \bottomrule
    \end{tabular}
    }
    \vspace{-1.2em}
\end{table}

\begin{table*}[t]
    \centering
    \caption{Validation (Val) and test set classification performance across task formulations, including the confusion matrix for the validation-selected classifier on AMD grading. For general-domain baselines, only the best classifiers are reported. Best and second-best performances are shown in \textbf{bold} and \textit{italic}, respectively.}
    \label{tab:results}
    \vspace{-0.3em}
    \includegraphics[width=\linewidth, trim={0 0.1cm 0 0}, clip]{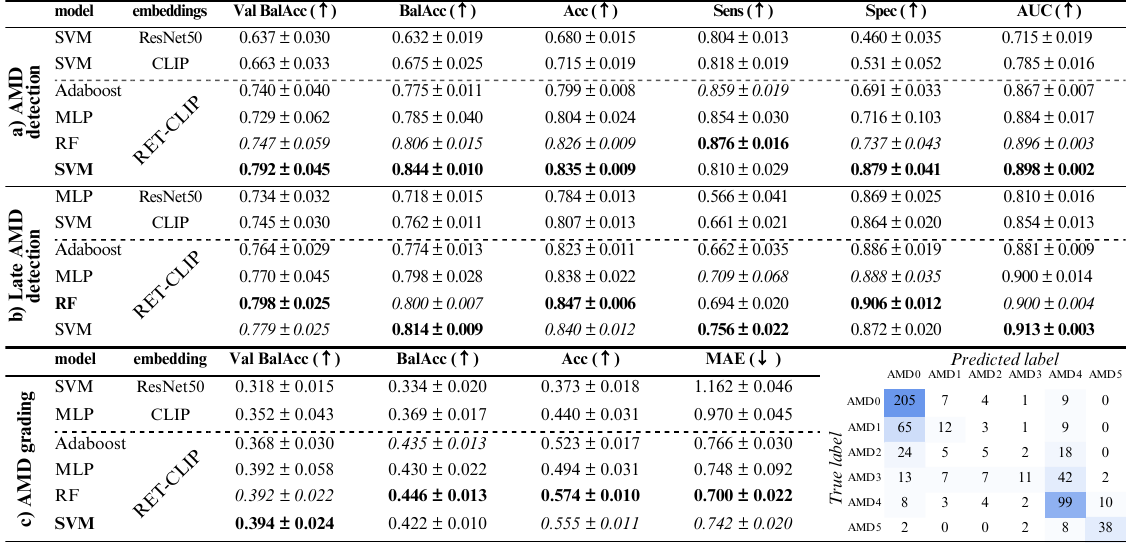}
    \vspace{-0.1em}
    {\footnotesize
    \textit{BalAcc}: Balanced Accuracy;
    \textit{Acc}: Accuracy;
    \textit{Sens}: Sensitivity;
    \textit{Spec}: Specificity;
    \textit{MAE}: Mean Absolute Error;
    \textit{AUC}: Area Under the Receiver Operating Characteristic Curve;
    \textit{MLP}: Multi-Layer Perceptrons;
    \textit{RF}: Random Forest;
    \textit{SVM}: Support Vector Machines.\par}
    \normalsize
    \vspace{-1.8em}
\end{table*}

Table~\ref{tab:results} summarizes the RET-CLIP based classification performance across the three task formulations.
For baselines, only the best validation-selected model is reported.
Across all classifiers on the test set, both AMD detection and Late-AMD detection achieved AUC above \(0.867\pm0.007\) and balacc above \(0.774\pm0.013\), while six-class grading reached a maximum balacc of \(0.446\pm0.013\), highlighting the strong influence given by the task definition. 
RET-CLIP embeddings consistently outperformed both general-domain baselines across all three tasks (e.g. AMD detection balacc: \(0.844\) vs. \(0.675\) for CLIP and \(0.632\) for ResNet50), supporting the contribution of retinal-domain pretraining.
Notably, the same progressive decline across task granularities was also observed in both baselines, proving that the challenge of fine-grained severity discrimination is not unique to RET-CLIP.
The low standard deviations observed across the ten independent repetitions further indicate that the reported trends are stable and reproducible.
Classifier choice significantly affected AMD detection: SVM differed significantly from every other model, and RF differed significantly from both MLP and AdaBoost (5 of 6 pairwise comparisons), with only MLP and AdaBoost performing comparably. 
In contrast, classifier selection had no significant effect on Late-AMD detection (0 of 6); only between SVM and RF the pairwise difference was significant.
Despite this variability in classifier-level significance, the performance variations remained substantially smaller than those observed across task formulations, suggesting that the organization of the RET-CLIP representation space plays a larger role in determining task difficulty.

Despite the reduced performances, a more detailed analysis of the AMD grading task revealed a marked asymmetry across classes.
Healthy eyes (AMD 0) and advanced disease stages (AMD 4--5) achieved substantially higher accuracies than intermediate stages.
As shown in the confusion matrix (Tab.~\ref{tab:results}), the validation-selected SVM classifier reached class accuracies of \(0.907\), \(0.786\), and \(0.760\) for AMD scores 0, 4, and 5, respectively, while scores 1--3 remained below \(0.15\), with most of their misclassifications falling into nearby severity levels.
This behavior closely mirrors the latent-space organization observed in Fig. \ref{fig:umap} and Fig. \ref{fig:distances}, where healthy and advanced AMD occupied more distinct regions of the RET-CLIP representation space, while intermediate stages exhibited substantial overlap and increased intra-class variability \cite{peng2019deepseenet}.

Together, these findings suggest that the difficulty of AMD severity assessment is influenced not only by the classification task itself, but also on how disease severity is represented within the learned feature space.
In particular, healthy and advanced AMD cases consistently formed more distinguishable representation patterns identified across all classifiers, whereas intermediate stages remained poorly separable regardless of the learning paradigm.
This observation is consistent with our representation-space analysis and reflects the RET-CLIP pretraining objective, which was optimized for contrastive vision-language alignment rather than fine-grained discrimination.

\begin{figure}[ht!] 
    \centering
    \includegraphics[width=\linewidth]{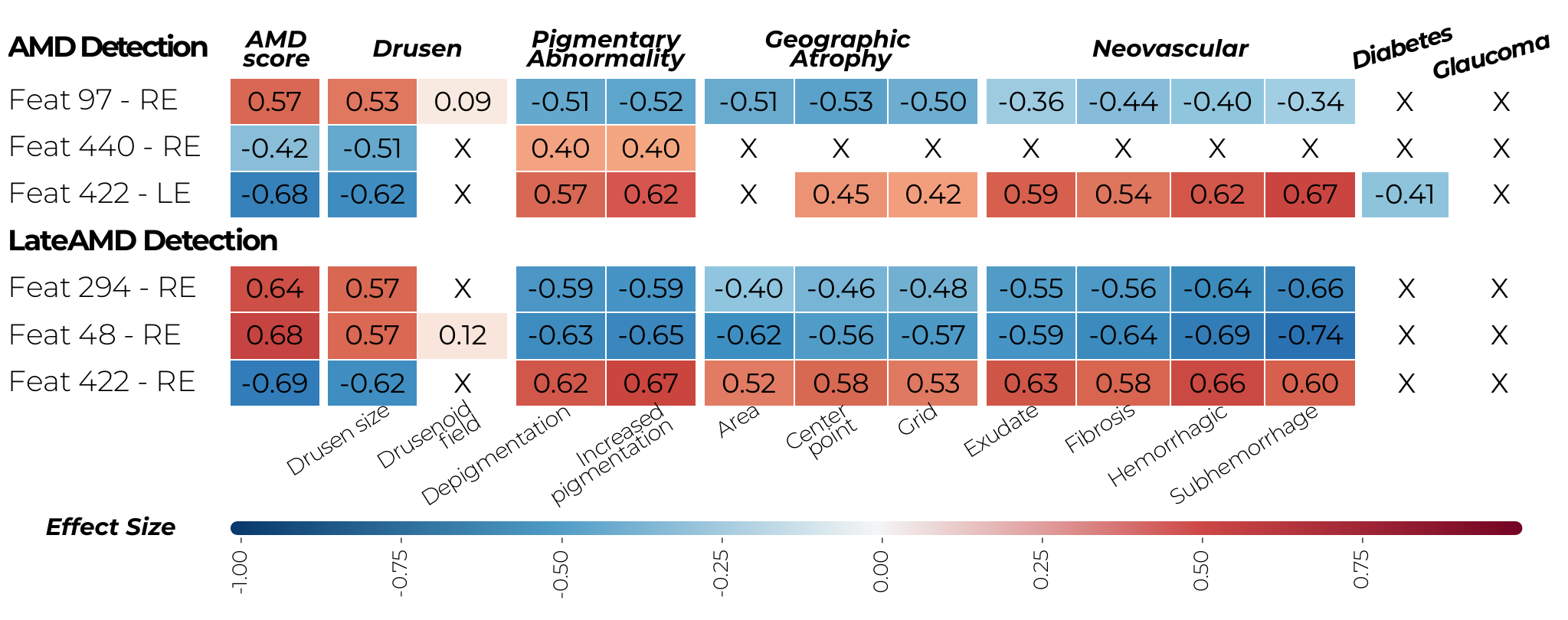} 
    \caption{Phenotype associations for SHAP-selected latent dimensions. Heatmap shows the association strength: Spearman's correlation coefficient for ordinal phenotypes, Brunner--Munzel rank-biserial correlation for binary phenotypes. Only associations statistically significant after patient-cluster bootstrap and BH-FDR correction reported. X marks non-significant pairs.}
    \label{fig:statistical}
    \vspace{-1.5em}
\end{figure}

Since the six-class grading performance was largely driven by the correct identification of healthy and late-AMD stages, we focused the subsequent interpretability analysis on the two binary classification tasks.
The lower performance may partly reflect a mismatch with AREDS staging, while core discriminative signals are already captured by the binary classifiers.
In particular, SVM was selected for the AMD detection (validation balacc \(0.792\pm 0.045 [0.748, 0.823]\), test balacc \(0.844\pm 0.010 [0.784, 0.898]\)), while RF was selected for Late-AMD detection (validation balacc \(0.798\pm0.025 [0.731, 0.824]\), test balacc \(0.800\pm 0.007 [0.717, 0.882]\)).

\subsection{Latent Feature Interpretation}

\begin{figure*}[!t] 
  \centering
  \includegraphics[width=.59\linewidth, valign=c]{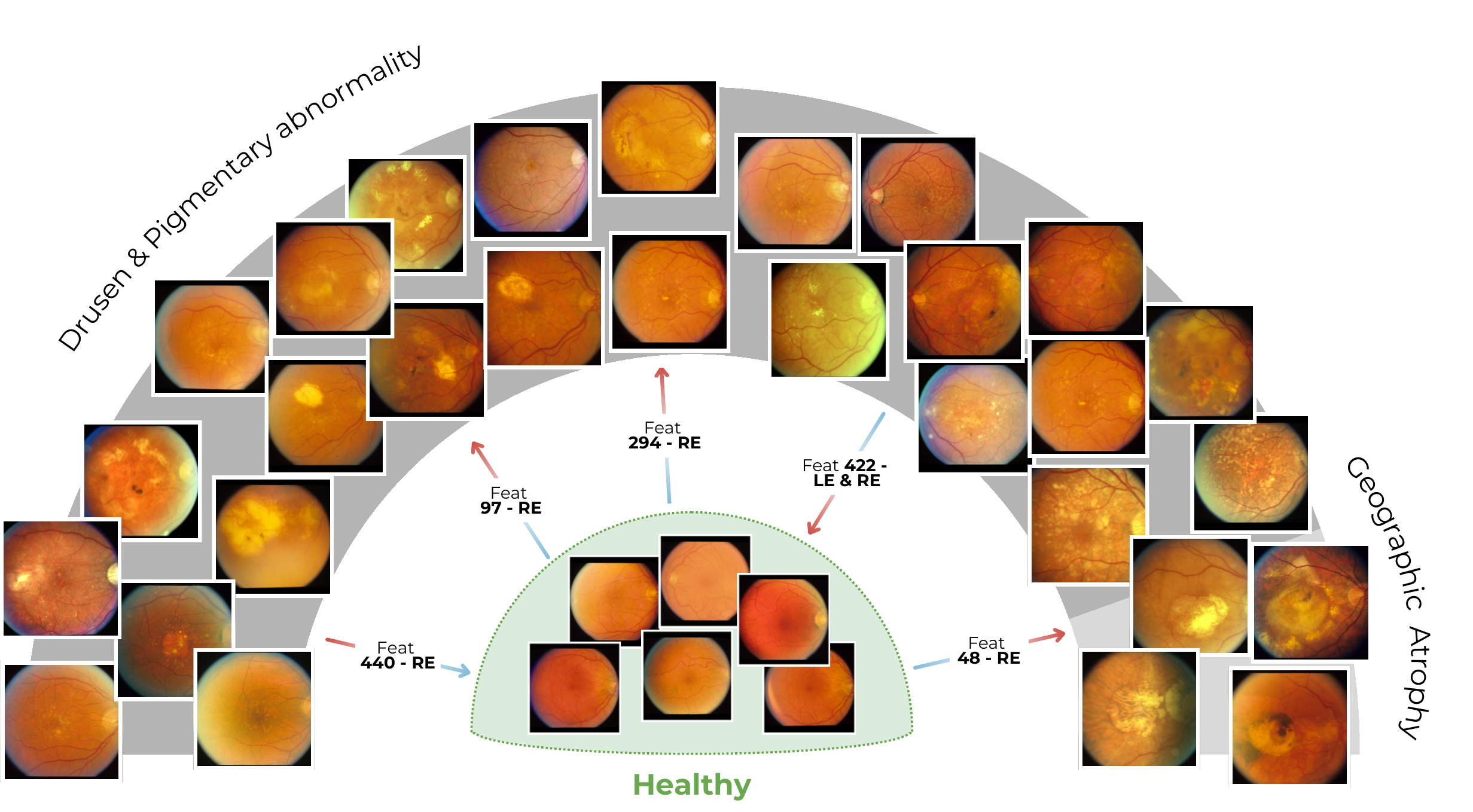}\hfill\vline\hfill
  \includegraphics[width=.40\linewidth, valign=c]{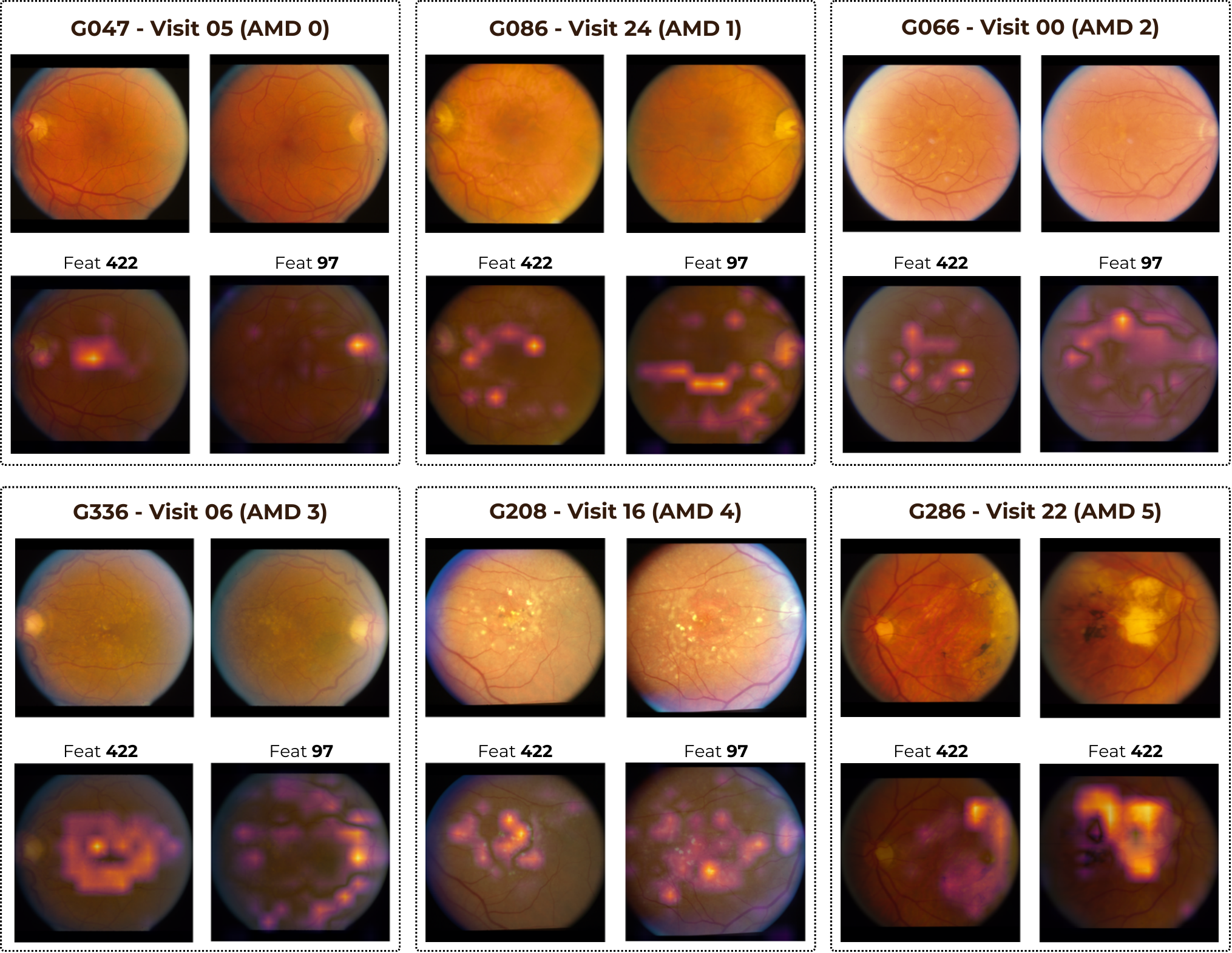}
  \caption{Qualitative interpretation of the most influential RET-CLIP latent dimensions. \textit{Left:} Representative high-activation fundus images for the selected latent dimensions, illustrating the retinal phenotypes captured by each feature. \textit{Right:} Representative latent-dimension localization maps highlighting the retinal regions in which the selected features are most strongly expressed across the AMD severity spectrum.} 
  \label{fig:qualitative}
  \vspace{-1.8em}
\end{figure*}

SHAP analysis identified a sparse subset of highly influential latent dimensions for each task (Fig. \ref{fig:statistical}).
Notably, most of them originated from the right-eye embedding, while Feature~422 emerged independently in both left- and right-eye representations, suggesting a consistent encoding of disease-related information across eyes. 
Further analysis on possible bias of the dataset or relevant patterns characterizing the right eye-only, such as variance between left- and right-eye representations, led to no relevant observations or additional assumptions extrapolable from the present data.
Although the selected features differed between the two clinical tasks, all exhibited strong and statistically significant associations with AMD score (\(p\text{-value}<0.001\) for all features), with effect size value ranging from \(-0.68\) to \(0.57\) and from \(-0.69\) to \(0.68\). 
Interestingly, the identified dimensions showed comparable correlation magnitudes but opposite directions, suggesting that RET-CLIP represents disease progression through complementary latent directions rather than a single disease-specific component.
Analysis of the association with retinal phenotypes (Fig. \ref{fig:statistical}) showed that the identified features were consistently associated with clinically relevant AMD biomarkers, particularly drusen size and pigmentary abnormalities.
Of note, Feature~440-RE showed no associations while Feature~48-RE strong correlation with geographic atrophy and neovascular, in line with the related task formulation.
In contrast, no significant associations were observed with glaucoma for any of the selected dimensions.
While one feature (422-LE) showed statistically significant associations with diabetes after FDR correction, the corresponding effect size was smaller than those observed for the AMD phenotypes.
These findings support the interpretation that the identified latent dimensions primarily encode AMD-related retinal characteristics rather than generic disease markers or task-specific statistical artifacts. 
Fig.~\ref{fig:qualitative}-\textit{left} shows representative high-activation images that consistently showed retinal abnormalities matching the phenotype associations identified statistically, while low-activation samples were predominantly healthy. 
Similarly, localization maps (Fig.~\ref{fig:qualitative}-\textit{right}) highlighted spatially coherent retinal regions corresponding to visible AMD-related abnormalities, with stronger and more extensive activations observed in advanced disease stages.
Although different latent dimensions exhibited distinct activation patterns, their localization frequently overlapped, suggesting that clinically relevant information is distributed across multiple latent dimensions.

Taken together, these results suggest that RET-CLIP does not encode AMD through a single disease-specific component.
Instead, AMD-related information appears to be distributed across multiple latent dimensions, each capturing distinct retinal phenotypes and complementary aspects of disease progression.
This observation further supports the value of analyzing foundation-model embedding spaces to better understand how clinically relevant information is encoded and utilized for downstream prediction tasks.

\subsection{Discussion}
In this study, we showed that RET-CLIP captures AMD-related information in a structured and clinically meaningful manner, representing severity as a continuous spectrum consistent with the clinical understanding of disease progression. 
This pattern persisted after accounting for within-class variability and class-distribution structure, supporting a graded representation with overlap between neighboring stages rather than sharply separated clusters.
The classification experiments suggested that limitations in AMD assessment arise not only from the predictive model itself but also from this intrinsic organization of disease severity, even as RET-CLIP's retinal pretraining showed a clear improvement over general-purpose embeddings on clinically meaningful downstream tasks.
Finally, the interpretability analysis demonstrated that RET-CLIP encodes complementary aspects of AMD pathology, with statistically significant associations with drusen accumulation, pigmentary abnormalities, and advanced retinal lesions, supporting the biological relevance of the learned representations.

While previous retinal foundation-model studies have primarily focused on downstream predictive performance \cite{jiao2025foundation}, comparatively little attention has been devoted to understanding how disease-related information is organized within the learned representation space.
Our findings contribute to this emerging research direction by linking representation-space structure, downstream task performance, and clinically meaningful retinal phenotypes.

When compared with previous AMD classification studies, our RET-CLIP-based framework achieved competitive performance for binary AMD detection, despite relying exclusively on frozen foundation-model embeddings rather than task-specific end-to-end optimization.
While several supervised approaches have reported higher accuracies for AMD and Late-AMD detection, reaching up to \(0.98\)\cite{xie_automation_2023} and \(0.91\)\cite{heo2020development,el2023scale} compared with \(0.835 \pm 0.009\) and \(0.847 \pm 0.006\) in the present study, our results demonstrate that RET-CLIP embeddings retain substantial AMD-related information and support clinically relevant disease discrimination with minimal task-specific adaptation.
A different trend was observed for AMD grading, where performance decreased substantially.
While dedicated supervised frameworks built on explicit biomarker integration yielded accuracies of \(0.67\)~\cite{peng2019deepseenet} and \(0.80\)~\cite{xie_automation_2023}, our goal was to investigate foundation model's latent disease organization rather than maximize grading performance.

Several limitations should be considered when interpreting these findings.
First, all experiments were performed using the AREDS cohort.
While AREDS is one of the largest and most extensively characterized AMD datasets available, external validation on independent cohorts will be necessary to assess the generalizability of the observed representation patterns.
Second, the latent-feature interpretation should be considered exploratory.
Although the identified dimensions were consistently associated with clinically meaningful retinal abnormalities and supported by qualitative ophthalmological assessment, only a limited subset of influential features was investigated. 
Third, although class imbalance was not explicitly addressed, it did not fully explain the performance differences, as some smaller advanced-stage classes were identified more accurately than larger intermediate ones.

Future studies should investigate whether similar representation patterns emerge across other retinal pathologies, such as diabetic retinopathy and glaucoma, and whether the identified latent dimensions can serve as prognostic biomarkers of future disease progression. 
Overall, our findings highlight the value of foundation-model representation analysis as a complementary research direction to downstream performance evaluation.
Understanding how clinically relevant information is organized within embedding spaces may support the development of more interpretable artificial intelligence based systems, facilitate the discovery of novel disease biomarkers, and provide new opportunities for disease progression modeling and prognosis estimation.

\section*{Conclusions}
This study investigated how AMD severity is represented within RET-CLIP foundation-model embedding space using bilateral retinal fundus images from the AREDS dataset. 
Through the analyses, we found that AMD-related information is not encoded as an arbitrary collection of predictive features, but rather as a structured representation that reflects clinically meaningful aspects of disease progression.
The observed organization of the latent space was consistent across experiments: it explained differences in classification performance across task formulations and revealed latent dimensions statistically associated with distinct retinal phenotypes, including drusen accumulation, pigmentary alterations, and advanced retinal lesions. 
Together, these findings suggest that analyzing the internal organization of retinal foundation-model representations can complement predictive evaluation, offering a new framework for improving interpretability and supporting future biomarker discovery and new prognostic modeling strategies.

\section*{Acknowledgment}\vspace{-0.1em}
This work was supported by World Premier International Research Center Initiative (WPI), MEXT, Japan.
Data was provided by the National Eye Institute (NEI) Age-Related Eye Disease Study (AREDS) repository (Application 36362).\vspace{.6em}

\bibliographystyle{IEEEbib}
\bibliography{bibliography}
\end{document}